\documentstyle[prl,multicol,aps,epsf]{revtex}

\begin{document}
\title{Embedded Solitons in a Three-Wave System}
\author{Alan R. Champneys \vspace{0.1cm}}
\address{Department of Engineering Mathematics, The University of Bristol,\\
Bristol BS8 1TR, United Kingdom \vspace{0.4cm} \\
Boris A.Malomed \vspace{0.1cm} \\
Department of Interdisciplinary Studies, Faculty of Engineering,\\
Tel Aviv University, Tel Aviv 69978, Israel}
\maketitle

\begin{abstract}
We report a rich spectrum of isolated solitons residing inside ({\it embedded%
} into) the continuous radiation spectrum in a simple model of three-wave
spatial interaction in a second-harmonic-generating planar optical waveguide
equipped with a quasi-one-dimensional Bragg grating. An infinite sequence of
fundamental embedded solitons are found, which differ by the number of
internal oscillations. Branches of these zero-walkoff spatial solitons give
rise, through bifurcations, to several secondary branches of walking
solitons. The structure of the bifurcating branches suggests a multistable
configuration of spatial optical solitons, which may find straightforward
applications for all-optical switching.
\end{abstract}


\begin{multicols}{2}
\narrowtext

\section{Introduction}

Recent studies have revealed a novel class of {\it embedded solitons} (ESs)
in various nonlinear-wave systems. An ES is a solitary wave which exists
despite having its internal frequency in resonance with linear (radiation)
waves. ESs may exist as {\it codimension-one} solutions, i.e., at discrete
values of the frequency, provided that the spectrum of the corresponding
linearized system has (at least) two branches, one corresponding to
exponentially localized solutions, the other one to delocalized radiation
modes. In such systems, quasilocalized solutions (or ``generalized solitary
waves'' \cite{Bo}) in the form of a solitary wave resting on top of a
small-amplitude continuous-wave (cw) background are generic \cite{KLM}.
However, at some special values of the internal frequency, the amplitude of
the background may exactly vanish, giving rise to an isolated soliton
embedded into the continuous spectrum.

Examples of ESs are available in water-wave models, taking into account
capillarity \cite{ChGr}, and in several nonlinear-optical systems, including
a Bragg grating incorporating wave-propagation terms \cite{we} and
second-harmonic generation in the presence of the self-defocusing Kerr
nonlinearity \cite{YKM} (the latter model with competing nonlinearities was
introduced earlier in a different context \cite{combined}).

It is relevant to stress that ESs, although they are isolated solutions, are 
{\em not} structurally unstable. Indeed, a small change of the model's
parameters will slightly change the location of ES (e.g., its energy and
momentum, see below), but will not destroy it, which is clearly demonstrated
by the already published results \cite{ChGr,YKM}. In this respect, they may
be called generic solutions of codimension one.

ESs are interesting because they naturally appear when higher-order
(singular) perturbations are added to the system, which may completely
change its soliton spectrum. Optical ESs have a potential for applications,
due to the very fact that they are isolated solitons, rather than occurring
in continuous families. The stability problem for ESs was solved in a fairly
general analytical form in Ref. \cite{YKM}, which was also verified by
direct simulations of the model considered. It was demonstrated that ES is a 
{\it semi}-stable object which is fully stable to linear approximation, but
is subject to a slowly growing (sub-exponential) one-sided nonlinear
instability. Development of this weak instability depends on values of the
system's parameters; in some cases, it is developing so slowly that ES, to
all practical purposes, may be regarded as a fully stable object 
\cite{unpub}.

In the previously studied models, only a few branches of ESs were found, and
only after careful numerical searching, which suggest they may be hard to
observe in a real experiment. The present work aims to investigate ESs in a
recently introduced model of a three-wave interaction in a quadratically
nonlinear planar waveguide with a quasi-one-dimensional Bragg grating \cite
{Mak}, which can be quite easily fabricated. It will be found that ESs occur
in abundance in this model, hence it may be much easier to observe them
experimentally. It should also be stressed that, unlike the previously
studied models, in which ESs appear in relatively exotic conditions, e.g.,
as a result of adding singular perturbations \cite{we} or specially
combining different nonlinearities \cite{YKM}, the model that will be
considered below and found to give rise to a rich variety of ESs, is exactly
the same which was known to support vast families of ordinary (non-embedded)
gap solitons. This, in particular, implies that ES can be found in the
corresponding system under the same conditions which are necessary for the
observation of the regular solitons, i.e., the experiment may be quite
straightforward. An estimate of the relevant physical parameters will be
given at the end of the paper.

The rest of the paper is organized as follows. In section 2, we recapitulate
the model and obtain solutions in the form of fundamental {\it zero-walkoff}
ESs, which, physically, correspond to the case when the Poynting vector of
the carrier waves is aligned with the propagation direction. The analysis is
extended in section 3 to the case of fundamental {\it walking} ESs, for
which the Poynting vector and the propagation distance are disaligned.
Concluding remarks are collected in section 4.

\section{The Model and Zero-Walkoff Solitons}

The model describes {\it spatial solitons} produced by the second-harmonic
generation (SHG) in a planar waveguide, in which two components of the
fundamental harmonic (FH), $v_{1}$ and $v_{2}$, are linearly coupled by the
Bragg reflection on a grating in the form of a system of scores parallel to
the propagation direction $z$ (for a more detailed description of the model
see \cite{Mak}): 
\begin{eqnarray}
i(v_{1})_{z}+i(v_{1})_{x}+v_{2}+v_{3}v_{2}^{\ast } &=&0,  \label{v1} \\
i(v_{2})_{z}-i(v_{2})_{x}+v_{1}+v_{3}v_{1}^{\ast } &=&0,  \label{v2} \\
2i(v_{3})_{z}-qv_{3}+D(v_{3})_{xx}+v_{1}v_{2} &=&0.  \label{v}
\end{eqnarray}
Here $v_{3}$ is the second-harmonic (SH) field, $x$ is a normalized
transverse coordinate, $q$ is a real phase-mismatch parameter, and $D$ is an
effective diffraction coefficient. The diffraction terms in the FH equations
(\ref{v1}) and (\ref{v2}) are neglected as they are much weaker than the
artificial diffraction induced by the Bragg scattering, while the SH wave,
propagating {\it parallel} to the grating, undergoes no reflection, hence
the diffraction term is kept in Eq. (\ref{v}).

Experimental techniques for generation and observation of spatial solitons
in planar waveguides are now well elaborated (\cite{spatial}), and the
waveguide carrying a set of parallel scores with a spacing commensurate to
the light wavelength (which is necessary to realize the resonant Bragg
scattering) can be easily fabricated. Therefore, the present system provides
a medium in which experimental observation of ESs is most plausible. As
mentioned above, the observation of ES in this system should be further
facilitated by the fact that it supports a multitude of distinct ES states,
see below.

Eqs. (1)--(3) have three dynamical invariants: the Hamiltonian, which will
not be used below, the energy flux (norm) 
\begin{equation}
E\equiv \int_{-\infty }^{+\infty }\left[
|v_{1}(x)|^{2}+|v_{2}(x)|^{2}+4|v_{3}|^{2}\right] dx\,,  \label{E}
\end{equation}
and the momentum, 
\begin{equation}
P\equiv i\int_{-\infty }^{+\infty }\left( (v_{1})_{x}^{\ast
}v_{1}+(v_{2})_{x}^{\ast }v_{2}+2(v_{3})_{x}^{\ast }v_{3}\right) dx\,.
\label{P}
\end{equation}
The norm played a crucial role in the analysis of the ES stability carried
out in \cite{YKM}.

\begin{figure}[thp]
\epsfxsize 7.5cm 
\centerline{\epsffile{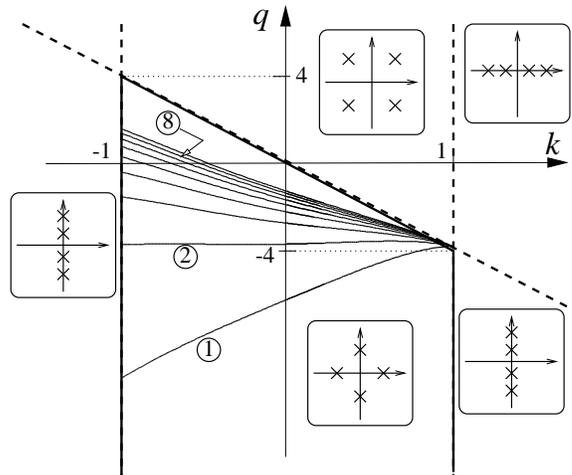}}
\caption{ The $(k,q)$ parameter plane of the three-wave model (\ref{v1}) -- (%
\ref{v}). The linear analysis (the results of which are summarized in the
inset boxes) shows that ES with $c=0$ may occur only in the region between
the solid bold lines. The bundle of curves emanating from the point $%
(k=1,q=-4)$ are branches of embedded-soliton solutions with $c=0$.}
\end{figure}

Soliton solutions to Eqs. (1)--(3) are sought in the form 
\begin{equation}
v_{1,2}(x,z)=\exp (ikz)\,u_{1,2}(\xi ),\;v_{3}(x,z)=\exp (2ikz)\,u_{3}\,,
\label{soliton}
\end{equation}
where $\xi \equiv x-cz$, with $c$ being the {\it walkoff} (slope) of the
spatial soliton's axis relative to the light propagation direction $z$. The
substitution of (\ref{soliton}) into Eqs. (1)--(3) leads to an 8th-order
system of ordinary differential equations (ODEs) for the real and imaginary
parts of $v_{1,2,3}$ (primes standing for $d/d\xi $): 
\begin{eqnarray}
-ku_{1}+i(1-c)u_{1}^{\prime }+u_{2}+u_{3}u_{2}^{\ast } &=&0,  \label{u1} \\
-ku_{2}-i(1+c)u_{2}^{\prime }+u_{1}+u_{3}u_{1}^{\ast } &=&0,  \label{u2} \\
-(4k+q)u_{3}+Du_{3}^{\prime \prime }-2icu_{3}^{\prime }+u_{1}u_{2} &=&0.
\label{u3}
\end{eqnarray}

Before looking for ES solutions to the full nonlinear equations, it is
necessary to investigate the eigenvalues $\lambda $ of their linearized
version, in order to isolate the region in which ESs {\it may} exist.
Substituting $u_{1},u_{2}\sim \exp (\lambda \xi )$, and $\,u_{3}\sim \exp
(2\lambda \xi )$ into Eqs. (6)--(8) and linearizing, one finds that the FH
and SH equations decouple in the linearized approximation. The FH equations
give rise to a biquadratic characteristic equation, 
\begin{equation}
(1-c^{2})^{2}\lambda ^{4}+2\left[ (1+c^{2})k^{2}-(1-c^{2})\right] \lambda
^{2}+(k^{2}-1)^{2}=0\,,  \label{4lambda}
\end{equation}
and the SH equation produces another four eigenvalues given by 
\begin{equation}
\left[ D\lambda ^{2}-(4k+q)\right] ^{2}+4c^{2}\lambda ^{2}=0.
\label{+4lambda}
\end{equation}

A necessary condition for the existence of ESs is that the eigenvalues given
by Eq. (\ref{4lambda}) have non-zero real parts - this is necessary for the
existence of exponentially localized solutions - while the eigenvalues from
Eq. (\ref{+4lambda}) should be purely imaginary (otherwise, one will have
ordinary, rather than embedded, solitons). This discrimination between the
two sets of the eigenvalues is due to the fact that Eqs. (\ref{u1}) and (\ref
{u2}) for the FH components are always linearizable, while the SH equation (%
\ref{u3}) may be {\it nonlinearizable}, which opens the possibility for the
existence of ESs \cite{YKM}. As it follows from Eqs. (\ref{4lambda}) and (%
\ref{+4lambda}), these two conditions imply 
\begin{equation}
k^{2}+c^{2}<1;\;4k+q<c^{2}/D\,.  \label{necessary}
\end{equation}
For the case $c=0$, the parametric region defined by the inequalities (\ref
{necessary}) is displayed in Fig. 1.

\begin{figure}[thp]
\epsfxsize 8.5cm
\centerline{\epsffile{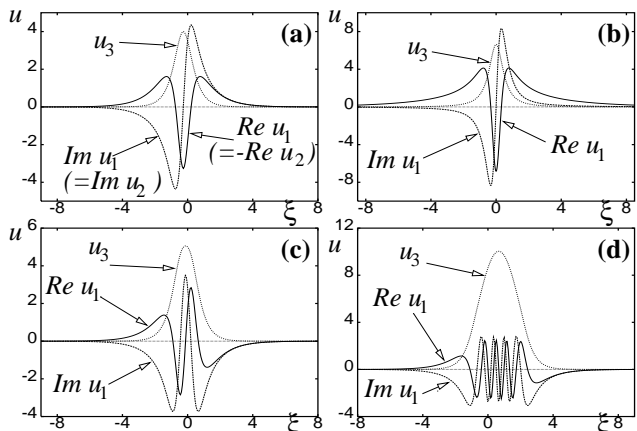}}
\caption{Typical examples of the fundamental embedded solitons with the zero
walkoff: (a) the ground-state for $k=0$; (b) the same solution for $k =-0.95$%
; (c,d) the first and eighth ``excited states" for $k=0$.}
\end{figure}

In Ref. \cite{Mak}, numerous ordinary ({\it gap} \cite{gap}) soliton
solutions to the present model have been found by means of a numerical
shooting method. To construct ES solutions, we applied the same method to
Eqs. (\ref{u1}), (\ref{u2}) and (\ref{u3}), allowing just one parameter to
vary. From each ES solution that was found this way, branches of the
solutions were continued in the parameters $k,\,q$ and $c$, by means of the
software package AUTO \cite{AUTO}. Note that the $c=0$ solutions admit an
invariant reduction $u_{2}=-u_{1}^{\ast }$, $\,u_{3}=u_{3}^{\ast }$, which
reduces the system to a 4th-order ODE system, thus making numerical shooting
feasible.

We confine the analysis to {\it fundamental} solitons, which implies that
the SH component $u_{3}$ has a single-humped shape (a distinctive feature of
gap solitons in the same system is that not only fundamental solitons, but
also certain double-humped two-solitons (bound states of two fundamental
solitons) appear to be stable \cite{Mak}). Note that double- and
multi-humped ESs must exist too as per a theorem from Ref. \cite{MHO}, but
leaving them aside, we will still find a rich structure of fundamental ESs.

\begin{figure}[thp]
\epsfxsize 6.5cm 
\centerline{\epsffile{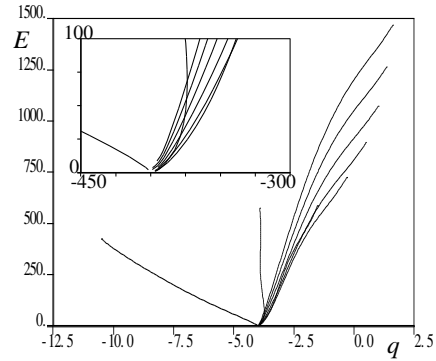}}
\caption{A diagram of the $c=0$ embedded solitons on the (energy-flux,
mismatch) plane. The inset zooms the most interesting part of the diagram.}
\end{figure}

We begin with a description of the results from the reduced case $c=0$, when
an additional scaling allows us to set $D\equiv 1$ without loss of
generality. The results are displayed in Figs.\ 1 -- 3. There is a strong
evidence for existence of an infinite ``fan'' of fundamental ES branches.
Among them, we define a {\it ground-state} soliton as the one which has the
simplest internal structure (Fig. 2a). The next ``first excited state''
differs by adding one (spatial) oscillation to the FH field (Fig. 2c).
Adding each time an extra oscillation, we obtain an indefinitely large
number of \ ``excited states'' (as an example, see the 8th state in Fig.
2d). We stress, however, that all the ``excited states'' belong to the class
of the fundamental solitons, rather than being bound states thereof.

In Fig. 1, the first nine states (branches) are shown in the $(k,q)$
parametric plane. Note that the whole bundle of the branches originates from
the point $(k=1,q=-4)$, which is precisely the intersection of the two lines
which limit the existence region of ES (see Eq. (\ref{necessary}) with $c=0$%
). At this degenerate point, the linearization (see above) gives four zero
eigenvalues. More branches than those depicted in Fig. 1 have been found,
the numerical results clearly pointing towards the existence of {\em %
infinitely many} branches, accumulating on the border $q+4k=0$ of the
ES-region. In the accumulation process, each $u_{3}$ component is
successively wider, while the $u_{1,2}$ ones have more and more internal
oscillations.

Since $k$ is an arbitrary propagation constant, from physical grounds, the
results obtained for the $c=0$ solutions are better summarized in terms of
energy flux $E$ vs. mismatch $q$\ (Fig. 3). Note that all the branches shown
in Fig. 3 really terminate at their edge points, which exactly correspond to
hitting the boundary $k=-1$, see Fig. 1. It is also noteworthy that all the
solutions are exponentially localized, except at the edge point $k=-1$,
where a straightforward consideration of Eqs. (\ref{u1})--(\ref{u3})
demonstrates that, in this case, ES are weakly (algebraically) localized as 
$|x|\rightarrow \infty $ (cf.\ Fig 2b): 
\[
\begin{array}{c}
u_{1}\approx \sqrt{-(4k+q)}|x|^{-1},\,u_{2}\approx (1/2)\sqrt{-(4k+q)}%
|x|^{-2}, \\ 
u_{3}\approx x^{-2}\,.
\end{array}
\]

Finally, we observe from Figs.\ 1 and 3 that the first ``excited-state''
branch has a remarkable property that it corresponds to a nearly constant
value of $q$. This means that while, generally, ES are isolated ({\it %
codimension-one}) solutions for fixed values of the physical parameters,
this branch is {\em nearly generic}, existing in a narrow interval of the $q$%
-values between $-4.0$ and $-3.74$.

\section{Walking Solitons}

We now turn to ESs with $c\neq 0$, i.e., {\em walking} ones. These were
sought for systematically by returning to the full 8th-order-ODE model and
allowing the AUTO package to detect bifurcations (of the {\it pitchfork}
type), while moving along branches of the $c=0$ solutions. It transpires
that {\em all} the bifurcating branches have $c\neq 0$, i.e., they are {\em %
walking} ESs. Such solutions are of {\em codimension-two} in the parameter
space (i.e., the solutions can be represented by curves $k(q)$, $c(q)$),
which can be established by a simple counting argument after noting that the
8th-order linear system has two pairs of pure imaginary eigenvalues.
Alternatively, the walking ESs can be represented, in terms of the energy
flux and momentum (see Eqs. (\ref{E}) and (\ref{P})), by curves $E(q)$ and $%
P(q)$. We present results only for the walking solutions which bifurcate
from the ground and first excited $c=0$ states, while other walking ESs can
also be readily found.

It was found that the ground-state branch has exactly two bifurcation
points, giving rise to two distinct walking-ES solution branches (up to a
symmetry). These new branches are shown, in terms of the most physically
representative $c(q)$ and $E(q)$ dependences, in Fig. 4. Note that they,
eventually, coalesce and disappear. As the inset to Fig. 4b shows, they
disappear via a tangent (fold, or saddle-node) bifurcation.

The first excited state has three bifurcation points. One of them gives rise
to a short branch of walking ESs that terminates, while two others appear to
extend to $q=-\infty $ (their ostensible ``merger'' in Fig. 5 is an artifact
of plotting). It is known that, in the large-mismatch limit $q\rightarrow
-\infty $, the present three-wave model with the quadratic nonlinearity goes
over into a modified Thirring model with cubic nonlinear terms \cite{Trillo}%
. This suggests that the latter model may also support ES. However,
consideration of this issue is beyond the scope of the present work.

\begin{figure}[thp]
\epsfxsize 7.5cm
\centerline{\epsffile{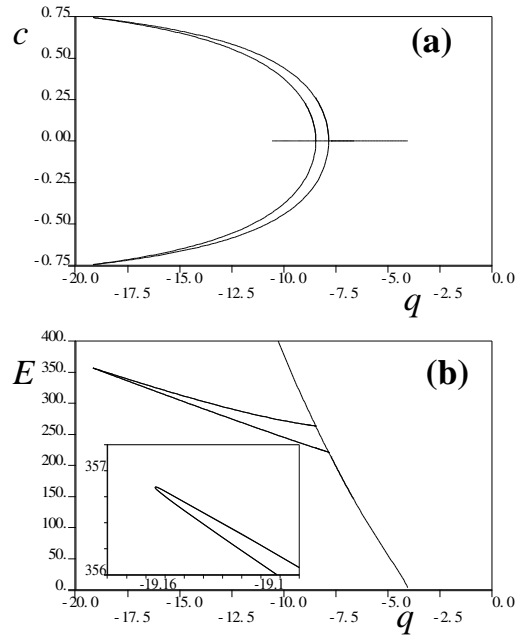}}
\caption{Two branches of ``walking" ($c\neq0$) embedded solitons bifurcating
from the ground-state $c=0$ branch: (a) the walkoff $c$, and (b) the energy
flux $E$ vs. the mismatch $q$. The horizontal segment in (a) shows the
branch of the $c=0$ solutions. The inset in (b) shows that the two branches
meet and disappear via a typical tangent bifurcation.}
\end{figure}

Fig. 4 clearly shows that, in a certain interval of the mismatch parameter $q
$, the system gives rise to a {\em multistability}, i.e.,\ coexistence of
different types of spatial solitons in the planar optical waveguide (for
instance, taking account of the fact that each $c\neq 0$ branch has
symmetric parts with the opposite values of $c$, we conclude that there are 
{\it five} coexisting solutions at $q$ taking values between about $-8$ and $%
-11$). This situation is of obvious interest for applications, especially in
terms of all-optical switching \cite{spatial}. Indeed, switching from a
state with a larger value of the energy flux to a neighboring one with a
smaller flux can be easily initiated by a small localized perturbation, in
view of the above-mentioned {\it one-sided} semistability of ES, shown in a
general form in \cite{YKM}. Such a switching perturbation can be readily
made controllable and movable if created by a laser beam launched normally
to the planar waveguide and focused at a necessary spot on its surface \cite
{Wang}. Switching between the two branches with $c\neq 0$ can be quite easy
to realized too, due to small energy-flux and walkoff/momentum differences
between them, see Fig. 4.

\section{Conclusion}

To conclude the analysis, it is necessary to estimate the actual size of the
relevant physical parameters. This is, in fact, quite easy to do, as there
is no essential difference in the estimate from that which was presented in
Ref. \cite{Mak} for the ordinary solitons in exactly the same model. This
means that a diffraction length $\sim 1$ cm is expected for the SH
component, and, definitely, the diffraction lengths for the FH components,
which are subject to the strong Bragg scattering, will be no larger than
that. Thus, a sample with a size of a few cm may be sufficient for the
experimental observation of ESs. The sample may be an ordinary planar
quadratically nonlinear waveguide with a set of parallel scores written on
it. The other parameters, such as the power of the laser beam that generates
the solitons, etc., are expected to be the same as in the usual experiments
with the spatial solitons \cite{spatial}. As concerns the weak
semi-instability of ESs, it may be of no practical consequence for the
experiment, as it would manifest itself only in a much larger sample. In
this connection, it may be relevant to mention that, strictly speaking, the
usual spatial solitons observed in numerous experiments are all unstable
(e.g., against transverse perturbations) in the usual (linear) sense, but
the instability has no room to develop in real experimental samples.

Finally, we see from Figs. 4 and 5 that the maximum walkoff that ESs can
achieve is, in the present notation, slightly smaller than $1$. According to
the geometric interpretation of the underlying equations (\ref{v1}) - (\ref
{v}) (see details in the original work \cite{Mak}), this implies that the
maximum size of the misalignment angle between the propagation direction and
the axis of the spatial soliton may be nearly the same as the (small) angle
between the Poynting vectors of the two FH waves and that of the SH wave.

To summarize the work, we have found a rich spectrum of isolated solitons
residing inside the continuous spectrum in a simple model of the three-wave
spatial interaction in a second-harmonic-generating planar optical waveguide
equipped with a quasi-one-dimensional Bragg grating. An infinite sequence of
fundamental embedded solitons were found. They differ by the number of
internal oscillations. The embedded solitons are localized exponentially,
except for a limiting degenerate case, when they become algebraically
localized. Branches of the zero-walkoff spatial solitons give rise, through
bifurcations, to several branches of walking solitons. The structure of the
bifurcating branches provides for a multistable configuration of the spatial
optical solitons. This may find straightforward applications to all-optical
switching.

\section*{Acknowledgements}

The stay of B.A.M. at the University of Bristol was supported by a Benjamin
Meaker fellowship. A.R.C. holds and U.K. EPSRC Advanced Fellowship.

\begin{figure}[thp]
\epsfxsize 7.5cm
\centerline{\epsffile{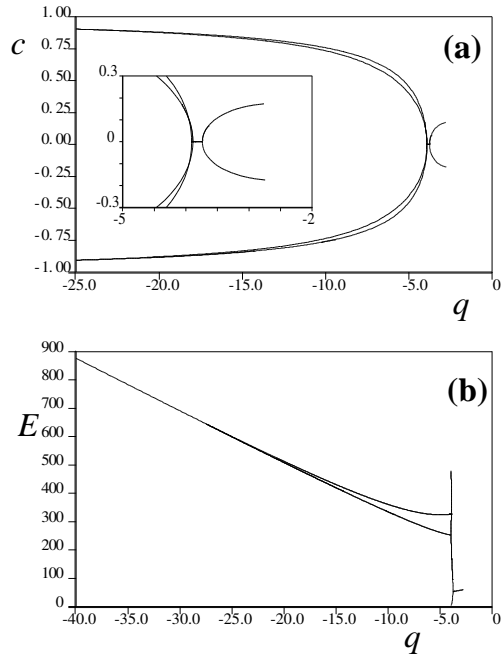}}
\caption{Three branches of the walking $(c\neq 0)$ embedded solitons
bifurcating from the $c=0$ branch corresponding to the first ``excited
state'', depicted similarly to Fig. 4 The inset in (a) shows in detail the
central part of the diagram.}
\end{figure}

%\section*{FIGURE CAPTIONS}

\end{multicols}

\end{document}